\documentclass[a4paper]{article}
\usepackage[]{geometry}
\usepackage[utf8]{inputenc}
\usepackage[T1]{fontenc}
\usepackage{graphicx}
\usepackage{scrhack}

\usepackage[usenames,dvipsnames,svgnames,table]{xcolor}
\definecolor{gris}{RGB}{200,200,200}
\usepackage{algorithm,algorithmic}
\usepackage{amsfonts}
\usepackage[fleqn]{amsmath}
\usepackage{amssymb}

\newcolumntype{B}[1]{>{\centering\arraybackslash}b{#1}}

\usepackage{hyperref}
\hypersetup{
  colorlinks=true,
  linkcolor=blue!50!red,
  urlcolor=green!70!black
}

\usepackage{listings}
\usepackage{fancyhdr}
\fancypagestyle{firststyle}
{
	\fancyhf{}
	\fancyfoot[L]{2016}
	\fancyfoot[C]{Fluent Data $\&$ Artfact}
	\fancyfoot[R]{\thepage}
}

\usepackage{authblk}

\title{MCMC Louvain for Online Community Detection}
\author[1,3]{Yves Darmaillac\thanks{yves.darmaillac@fluentdata.info}}
\author[2,4]{Sébastien Loustau\thanks{artfact64@gmail.com}}
\affil[1]{\small Laboratoire de Mathématiques et de leurs Applications - UMR CNRS 5142, Avenue
de l’Université, 64013 Pau cedex, France}
\affil[3]{Fluent Data\\9 Avenue Pasteur\\
64260 Louvie Juzon\\
Twitter : @fluentdata}
\affil[2]{Laboratoire Angevin de Recherche en Mathématiques - UMR CNRS 6093, 2 Boulevard Lavoisier, 49045 Angers Cedex}
\affil[4]{Artfact, Technopole Hélioparc\\2 Avenue du President Pierre Angot\\
64000 Pau\\Twitter : @artfact64}

\newcommand{\bc}{C}
\newcommand{\R}{\mathbb{R}}
\newcommand{\1}{\mathbf{1}}
\newcommand{\C}{\mathcal{C}}
\newcommand{\Front}{\mathcal{F}}
\newcommand{\PP}{\mathcal{P}}
\newcommand{\V}{\mathcal{N}}
\newcommand{\map}{\text{map}}
\begin{document}
	
	\maketitle
	\thispagestyle{firststyle}
	\bigskip
	\abstract{We introduce a novel algorithm of community detection that maintains dynamically a community structure of a large network that evolves with time. The algorithm maximizes the modularity index thanks to the construction of a randomized hierarchical clustering based on a Monte Carlo Markov Chain (MCMC) method. Interestingly, it could be seen as a dynamization of Louvain algorithm (see \cite{blondel08}) where the aggregation step is replaced by the hierarchical instrumental probability.
}
	\tableofcontents

\section{Introduction}
Community detection has become very popular in network analysis the last decades. Its range of applications include social sciences, biology and complex systems, such as the world-wide-web, protein-protein interactions, or social networks (see \cite{fortunato} for a thorough exposition of the topic). To tackle this problem, spectral approaches have been introduced in \cite{MR2282139} or \cite{White05aspectral}, inspired from the so-called spectral clustering problem (see \cite{spectralclustering}). However, the treatment of larger and larger graphs has been investigated and modularity-based algorithms has been proposed. This class of algorithms maximize a quality index called modularity, introduced in \cite{newman}. Unfortunately, exact modularity optimization is NP-hard (see \cite{Brandes}) and becomes computationaly intractable for large networks. With this in mind, approximated solutions based on greedy search has been introduced, such as for instance \cite{Clauset}, or more recently \cite{blondel08}. For a complete comparison of recent methods against a benchmarks of graphs, we refer to \cite{comparativeanalysis}.
\medskip

In this paper, we introduce a modularity-based algorithm that provides a clustering of a dynamic graph. Dynamic graph clustering is a difficult problem which consists in maintaining dynamically a community structure such that at any time, there is higher density of edges whithin groups than between them. As far as we know, this problem has been poorly treaten in the literature. \cite{maillard} is the most serious attempt (see also the references therein), which provides dynamization of greedy search algorithms introduced in \cite{Clauset,blondel08}. In the supervised case, \cite{rakhlintuto} investigate online nodes classification where labels are correlated with the graph structure. It leads to algorithms based on penalized empirical risks and gradient computations.
\medskip

The philosophy behind this paper is to maintain dynamically a Markov Chain of hierarchical graphs and partitions that optimize the modularity index. For this purpose, we use the Metropolis Hasting (MH) algorithm, named after \cite{metropolis53} and \cite{hastings70}. MH algorithms is the simplest and more versatile solution to construct a Markov chain associated with a stationary distribution (see \cite{robertmh} for a self-contained introduction with R codes). MH is one of the most general MCMC algorithms. Given a target density probability $f$, it requires a working conditional probability $q$ also called proposal. The transition from the value of the Markov chain at time $k$ and its value at time $t+1$ proceeds via 2 steps : generate a proposal with law $q$, and accept this proposal with a suitable chosen acceptance ratio. The computation of this ratio preserves the stationary density $f$ without any assumption, and do not depend on the proposal distribution $q$. However, in practice, the performance of the algorithm strongly depends on the choice of the transition $q$, the real issue of MCMC algorithms since some choices see the chain unables to convergence in a reasonable time.
\medskip

The paper is organized as follows. In Section \ref{sec:2}, we introduce the first notations and describe the modularity-based algorithm proposed in \cite{blondel08}. A first static Metropolis-Hastings (MH) algorithm is derived. It gives similar results than in \cite{blondel08}, where each node is visited several times. In Section \ref{sec:3} and Section \ref{sec:4}, we propose improved versions of this static MH algorithm which combines local changes and includes aggregation, in order to speed up convergence. These considerations allow to construct a new competitive algorithm for static community detection. Section \ref{sec:5} is dedicated to the dynamic version of this algorithm, where we observe a sequence of graphs that evolves with time. Section \ref{sec:6} concludes the paper with a short discussion.
\section{Notations and preliminary study}
\label{sec:2}
\subsection{Notations}
Let $G=(V,E)$ an undirected and -possibly- weighted graph where $V$ is the set of $N$ vertices or nodes and $E$ the set of edges $(i,j)$, for $i,j\in\{1, \ldots, N\}$. We denote by $A\in\mathcal{M}_N(\R)$ the corresponding symmetric adjacency matrix where entry $A_{ij}$ denotes the weight assigned to edge $(i,j)$. The degree of a node $i$ is denoted $k_i$ and $m:=|E|=\frac{1}{2}\sum_i k_i$. We call $\bc\in \C$ a \textit{coloration} of graph $(V,E)$ any partition $\bc=\{c_1,\ldots,c_k\}$ of $V$ where for any $i=1,\ldots, k$, $c_i\subseteq V$ is a set of nodes of $G$. Moreover, with a slight abuse of notation, $\bc(i)\in\{1, \ldots, k\}$ denotes the community of vertex $i$ based on partition $\bc$.

With thess notations, the modularity $\bc\mapsto Q^{\bc}$ of a given graph $(V,E)$ is given by :
\begin{align}
\label{modularity}
Q^{\bc} = \frac{1}{2m}\sum_{i,j \in V^2} \left(A_{ij} - \frac{k_i k_j}{2m}\right)\delta\big(\bc(i),\bc(j)\big)
\end{align} where $\delta$ is the Kronecker delta. Roughly speaking, modularity compares fraction of edges that falls into communities of $\bc$ with its expected counterpart, given a purely random rewiring of edges which respect to nodes degrees $(k_i)_{i\in V}$. Maximization of \eqref{modularity} is NP-hard (see \cite{Brandes}), and heuristic approximation such as greedy search may suffer from local optima. However, the variation of modularity induced by local moove (such as moving an isolated node into an existing community, or remove one node in an existing community to a single node community) can be easily computed.
\medskip

This fact is at the core of Louvain algorithm (see \cite{blondel08}) and provides very fast graph clustering method. This algorithm iterates two phases : an optimization phase lets each node moving to one of its neighbors'clusters, in order to maximize the modularity index, whereas in the aggregation phase, each cluster is contracted to one node and edges weights are summed. These two phases are iterated several times until a stop criterion, and reveal a hierarchical structure usefull in practice where natural organization are observed. 
\subsection{Metropolis Hasting Algorithm}
In this subsection, instead of choosing the local moove which maximizes the modularity gain as in \cite{blondel08}, we use the MH algorithm described in \autoref{algomh} below.

\begin{algorithm}
\caption{MH for Community Detection}
\label{algomh}
\begin{algorithmic}[1]
  \scriptsize
  \STATE Initialization $\lambda>0$, $\bc^{(0)}$.
  \STATE For $k=1, \ldots, N$:
  \STATE Draw $\bc'\sim p(\cdot|\bc^{(k-1)})$ where $p(\cdot|\bc^{(k-1)})\in\PP(\V^{\bc^{(k-1)}})$ is the proposal distribution over $\V^{\bc^{(k-1)}}$, a neighborhood of $\bc^{(k-1)}$.
  \STATE Update $\bc^{(k)}=\bc'$ with acceptance ratio :
\begin{align}
\label{ratio}
\rho = 1\wedge \left(r_{\bc^{(k-1)} \rightarrow \bc'}\frac{\exp\left(\lambda Q^{\bc'}\right)}{\exp\left(\lambda Q^{\bc^{(k-1)}}\right)}\right),\mbox{ where }r_{\bc \rightarrow \bc'}:=p(\bc^{(k-1)}|\bc')/p(\bc'|\bc^{(k-1)}).
\end{align}
\end{algorithmic}
\end{algorithm}

\autoref{algomh} above satisfies the so-called detailed balance condition for any proposal $p$ and then produces a Markov chain with invariant probability density $f$ such that:
$$
f(\bc)d\bc\approx \exp\left(\lambda Q^{\bc}\right)d\bc.
$$
The major issue is then to define a particular neighborhood $\V^{\bc}$ and an idoine proposal $p(\cdot|\bc)$ in order to achieve convergence in a manageable time.

In this section, we propose a first attempt where neighborhood are local moove inspired from \cite{blondel08} whereas the proposal distribution takes advantages of the edges structure of the observed graph.
\paragraph{Neighborhood definition} In what follows, given $\bc\in\C$, the neighborhood $\V^{\bc}$ consists of all coloration $\bc'$ equals to $\bc$ except for one node $i\in V$. Then two cases arises: 
\begin{itemize}
\item $i$ joins an existing community $c\in\bc$ such that $c\not=\bc(i)$,
\item a new single node community $c_{\mathrm{new}}$ is created by $i$.
\end{itemize}
\paragraph{Proposal distribution}
The construction of the proposal distribution $p(\cdot|\bc)\in\PP(\V^{\bc})$ is based on two random choices : the choice of a node $i\in V$ and the choice of a community $c\in\bc$ thanks to an application $\Phi^{\bc}:V\times \bc\to \V^{\bc}$ such that $\bc'=\Phi^{\bc}(i,c)$ means that:
\begin{itemize}
\item $i$ joins an existing community $c$ if $(i,c)$ is such that $c\not=\bc(i)$,
\item a new single node community is created by node $i$ if $(i,c)$ is such that $c=\bc(i)$.
\end{itemize}
In \autoref{algomh}, the proposal distribution is then based on the previous mapping as follows:
\begin{itemize}
\item We first choose a node $i$ with discrete uniform probability over the set of nodes $V$;
\item Then we choose $c=C(j)$ where $j$ is chosen with law proportional to $A_{ij}$, excluding the case $j=i$.
\end{itemize}
To derive $\bc'$, we use the application $\Phi^{\bc}$ and state $$\bc'=\Phi^{\bc}(i,\bc(j)).$$

It is important to note that we exclude $j=i$ above in order to avoid identity move in \autoref{algomh}. Indeed, if $i=j$ and $i$ is a single node community, then $\bc'(i)=C(i)$ and $\bc'=\bc$.

We are now on time to define properly the proposal distribution. Let us fixed a coloration $\bc\in\C$. The proposal distribution is defined as follows:
\begin{align}
\label{ratio_num_1}
p(\bc'|\bc)=\begin{cases}
\dfrac{k^{\bc}_{i,\bc(i)}-A_{ii}}{(k_i-A_{ii})N} & \text{if $\bc'=\Phi^{\bc}(i,\bc(i))$,}\\[1em]
\dfrac{k^{\bc}_{i,c}}{(k_i-A_{ii})N}& \text{otherwise,}
\end{cases}
\end{align}
where for any node $i$, we denote by $k^{\bc}_{i,c}$  the total weight of edges from node $i$ to community $c$ as follows:
$$
k^{\bc}_{i,c}:=\sum_{j\in V}A_{ij}\mathbf{1}_{\bc(j)=c}.
$$
\paragraph{Acceptance ratio}
To compute the acceptance ratio in \eqref{ratio}, we need to calculate the probability $p(\bc|\bc')$ to come back. We have:
\begin{align}
\label{ratio_den_1}
p(\bc|\bc')=\begin{cases}
\dfrac{k^{\bc}_{i , \bc'(i)}}{(k_i-A_{ii})N}  & \text{if $\bc=\Phi^{\bc'}(i,\bc'(i))$,}\\[1em]
\dfrac{k^{\bc}_{i , \bc(i)}-A_{ii}}{(k_i-A_{ii})N} & \text{otherwise}
\end{cases}
\end{align}
The first case corresponds to the situation where $i$ was initially isolated in a single node community of $\bc$. 

Thus, the quantity $r_{\bc \rightarrow \bc'}$ in \eqref{ratio} is given by dividing \eqref{ratio_num_1} by \eqref{ratio_den_1}:
$$ r_{\bc \rightarrow \bc'} = \frac{p(\bc|\bc')}{p(\bc'|\bc)}=\begin{cases}
1& \text{if $\bc'=\Phi^{\bc}(i,\bc(i))$ or $\bc=\Phi^{\bc'}(i,\bc'(i))$,}\\[1em]
\dfrac{k^{\bc}_{i , \bc(i)}-A_{ii}}{k^{\bc}_{i,c}} & \text{otherwise.}
\end{cases}$$
\paragraph{Change of modularity}
Last step is to compute the likelihood in \eqref{ratio}. For this purpose, we introduce the quantity: 
$$\Delta Q^{\bc \rightarrow \bc'}:=Q^{\bc'}-Q^{\bc}.$$
It's easy to see from \eqref{modularity} that when an isolated node $i$ joins an existing community $c\in\bc$, we have:
\begin{align}
\Delta Q^{\bc \rightarrow \bc'} = \Delta Q^{\bc \rightarrow \bc'}_+:=\frac{1}{m}\sum_{j \in V}\left( A_{ij} - \frac{k_i k_j}{2m}\right)\delta(\bc(j),c) = \frac{1}{m}\left(k^{\bc}_{i,c} - \frac{ k_i k^{\bc}_{c}}{2m},\right) \label{eq:delta_plus}
\end{align} where $k^{\bc}_{c}=\sum_{j \in V}k_j \delta(\bc(j),c)$ is the total weight of community $c \in \bc$.

Symmetrically, when a node $i$ leaves its community $\bc(i)$ to form a new single node community:
\begin{align}
\Delta Q^{\bc \rightarrow \bc'}=\Delta Q^{\bc \rightarrow \bc'}_- :&= -\frac{1}{m}\sum_{j \in V}\left( A_{ij} - \frac{k_i k_j}{2m}\right)\delta(\bc(j),\bc(i)) \nonumber \\&= -\frac{1}{m}\left(k^{\bc}_{i,\bc(i)} - A_{ii}-\frac{ k_i }{2m}(k^{\bc}_{\bc(i)}-k_i)\right)\label{eq:delta_minus}
\end{align} 

Note that in \eqref{modularity} every term is summed twice due to the symmetry of the adjacency matrix. This explains the $1/m$ factor instead of $1/2m$ in \eqref{eq:delta_plus} and \eqref{eq:delta_minus}.

Next, the change of modularity incurred by our proposal is :

\begin{eqnarray}
\label{deltamodularity}
\Delta Q^{\bc \rightarrow \bc'} = \begin{cases}
\Delta Q^{\bc \rightarrow \bc'}_- & \text{if $\bc'=\Phi^{\bc}(i,\bc(i))$} \\
\Delta Q^{\bc \rightarrow \bc'}_- + \Delta Q^{\bc \rightarrow \bc'}_+ & \text{otherwise,} \\
\end{cases}
\end{eqnarray}
where we use in \eqref{deltamodularity} the fact that if $i$ joins an existing community, the change of modularity is equivalent to adding a new single node community with $i$ and then move this single node community to an existing community. This process has been introduced in \cite{blondel08} and allows a faster computation and the treatment of large graphs.


\subsection{Results}

\autoref{algomh} with previous proposal and acceptance ratio achieves the same kind of results than Louvain, where the number of proposals is roughly equals to the total number of inner loop iterations ( by testing modularity gain for each edge) in \cite{blondel08}. The modularity gain is slightly less and the number of communities is also slightly less than Louvain.

For instance, we have tested our algorithm on arxiv data and obtained an average of 55 communities and a modularity average of 0.810678. Standard Louvain (\cite{blondel08}) achieves slightly better results with an average of 58 communities and a modularity average of 0.820934.
\section{Improved proposal}
\label{sec:3}
\begin{figure}
	\centering
	\includegraphics[scale=0.5]{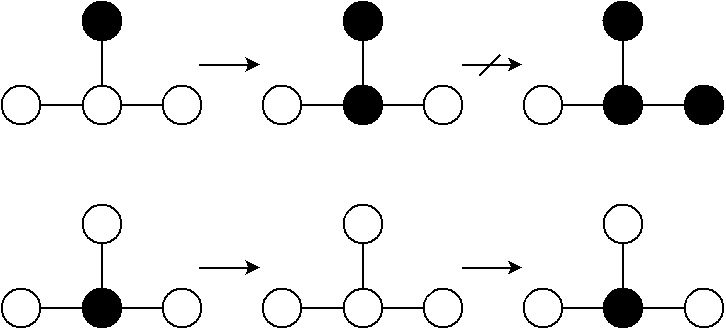}
	\caption{Two serious drawbacks}
	\label{fig:drawback}
\end{figure}

Actually the prior proposed in the previous section has two serious drawbacks (see figure \ref{fig:drawback}) : 

\begin{itemize}
	\item in the situation described by figure \ref{fig:drawback} upper part, the second move will never be accepted because the back proposal has a probability of zero,
	\item as communities grow and become bigger, more propositions of new single node communities that decrease modularity will occur, as shown in figure \ref{fig:drawback} lower part.
\end{itemize}

These considerations show slow convergence of the Markov chain and motivate the introduction of a better instrumental distribution. Indeed, to overcome the first drawback, the proposal should be able to allow a node to join a community potentially far from it. To overcome the second one, it should have higher probability to explores node at the boundary of the current state $\bc$. For that purpose, given $\bc\in\mathcal{C}$, we define a subset of $V$ made of nodes that have at least one edge in another community as follows:
$$
\Front^{\bc}:=\{i\in V:\exists j\in V \text{ such that } A_{ij}>0 \text{ and } \bc(j)\not=\bc(i)\}.
$$
This subset of $V$ is of particular interest and our proposition is to use a proposal distribution based on a mixture of a fully randomized distribution and a well-chosen distribution over the set $\Front^{\bc}$.
\subsection{Proposal definition}
The prior $p(\cdot|\bc)$ is defined as :
\begin{eqnarray}
\label{improvedproposal}
p(\cdot|\bc) = \alpha p_1(\cdot|\bc)+ \left(1-\alpha\right) p_2(\cdot|\bc),
\end{eqnarray}
where $\alpha\in(0,1)$ and $p_1(\cdot|\bc)$ and $p_2(\cdot|\bc)$ are defined as follows :
\begin{enumerate}
\item $p_1(\cdot|\bc)$ is equivalent to draw a first node $i$ uniformly over $V$ and a second one $j$ uniformly among the others,
\item $p_2(\cdot|\bc)$ is equivalent to draw a first node $i$ uniformly over $\Front^{\bc}$ and a second one $j$ proportionally to $k^\bc_{i,\bc(j)}$ with the constraint $\bc(i) \neq \bc(j)$.
\end{enumerate}
In both cases, to derive $\bc'$, we use the mapping $\Phi^{\bc}$ and state $\bc'=\Phi^{\bc}(i,\bc(j)).$

In the sequel we denote by  $|c|$ the cardinality of community $c$ based on coloration $\bc$ (we omit the dependence in $\bc$ for ease of exposition). Then, for any $i \in V$ and any $c \in \bc$ we have for $p_1$ :
$$
p_1(\bc'|\bc)=
\begin{cases}
\dfrac{|C(i)|-1}{N (N-1)} & \text{if $\bc'=\Phi^{\bc}(i,\bc(i))$,}\\[1em]
\dfrac{|c|}{N (N-1)} & \text{if $\bc'=\Phi^{\bc}(i,c)$ for $c\not=\bc(i)$,}
\end{cases}
$$
whereas for $p_2$:
$$
p_2(\bc'|\bc)=\begin{cases}\dfrac{k^\bc_{i,c}}{|\Front^{\bc}|\mathcal{K}^{\bc}(i)} & \text{if $\mathcal{K}^{\bc}(i) > 0$ and $c \ne \bc(i)$} \\[1em]
0 & \text{ otherwise}\end{cases}
$$
where $\mathcal{K}^{\bc}(i) = \sum_{\tilde{c}\in\bc} k^\bc_{i,\tilde{c}}\1_{\bc(i)\ne\tilde{c}}$. 

The addition of $p_1$ in \eqref{improvedproposal} allows to construct a better mixing strategy and derive a more efficient MH algorithm. Last step is to compute the conditional probabilities.  
\subsection{Conditional probabilities}
\label{condproba}
We will now compute $p(\bc'|\bc)$ and $p(\bc|\bc')$ from \eqref{improvedproposal}. Given node $i \in V$ and a coloration $\bc\in\C$, let us fix a coloration $\bc'\in\C$ and write $c=\bc'(i)$.  Then we have:
\begin{enumerate}
	\item If $\bc(i)$ and $c$ are not single node communities:
	\begin{align}
	\begin{split}
		&p\left(\bc'\mid\bc\right) = \frac{\alpha |c|^{\bc}}{N(N-1)}+\frac{(1-\alpha)k^\bc_{i,c}}{|\Front^{\bc}|\mathcal{K}^{\bc}(i)}, \\
		&p\left(\bc\mid\bc'\right) = \frac{\alpha |\bc(i)|^{\bc'}}{N(N-1)}+\frac{(1-\alpha)k^{\bc'}_{i,\bc(i)}}{|\Front^{\bc}|\mathcal{K}^{\bc'}(i)}= \frac{\alpha |\bc(i)|^{\bc}-1}{N(N-1)}+\frac{(1-\alpha)(k^{\bc}_{i,\bc(i)}-A_{ii})}{|\Front^{\bc}|(k_i-k^\bc_{i,c}-A_{ii})}.
	\end{split} \label{eq:ratio_mp}
	\end{align}

	\item If $c$ is a single node community :
	\begin{align}
	\begin{split}
		&p\left(\bc'\mid\bc\right) =\frac{\alpha (|\bc(i)|^{\bc}-1)}{N(N-1)}, \\
		&p\left(\bc\mid\bc'\right) = \frac{\alpha |\bc(i)|^{\bc'}}{N(N-1)}+\frac{(1-\alpha)k^{\bc'}_{i,\bc(i)}}{|\Front^{\bc}|\mathcal{K}^{\bc'}(i)}=\frac{\alpha |\bc(i)|^{\bc}-1}{N(N-1)}+\frac{(1-\alpha)(k^{\bc}_{i,\bc(i)}-A_{ii})}{|\Front^{\bc}|(k_i-A_{ii})}.
	\end{split} \label{eq:ratio_pp}
	\end{align}
	
	\item If $\bc(i)$ is a single node community :
	\begin{align}
	\begin{split}
			&p\left(\bc'\mid\bc\right) = \frac{\alpha |c|^{\bc}}{N(N-1)}+\frac{(1-\alpha)k^\bc_{i,c}}{|\Front^{\bc}|\mathcal{K}^{\bc}(i)}, \\
			&p\left(\bc\mid\bc'\right) = \frac{\alpha (|c|^{\bc'}-1)}{N(N-1)}=\frac{\alpha |c|^{\bc}}{N(N-1)}.
	\end{split} \label{eq:ratio_mm}
	\end{align}
\end{enumerate}
Finally we are able to compute $ r_{\bc \rightarrow \bc'}$ in \eqref{ratio}.

\section{Aggregation}
\label{sec:4}
Aggregation may accelerate convergence when communities are "big", i.e. when there is notably less communities than nodes. Aggregation consists of building a new graph which nodes are the communities of the former graph. This principle is adopted in \cite{blondel08} in the second phase of each pass after the optimization step as described in Section \ref{sec:2}. 
\medskip

In what follows, we propose to construct a MH type algorithm as in Section \ref{sec:3}, where aggregation is included in the proposal distribution in order to accelerate convergence. 
\subsection{Metropolis Hasting Algorithm with Hierarchical prior}
In our framework, we propose to use the same kind of prior than \eqref{improvedproposal} to a family of aggregated graphs in order to move entire communities rather than a single node. To achieve this, we introduce a set of hierarchical graphs and hierarchical priors as follows.

Let $G=(V,E)$ an undirected and -possibly- weighted graph where $V=\{1,\ldots, N\}$ is the set of $N$ vertices or nodes and $E$ is the set of edges $(i,j)$, for $i,j\in\{1, \ldots, N\}$. Let $L\geq 1$ an integer. The construction of a family of aggregated graphs $(E_l,V_l)_{l=1}^L$ is done iteratively as follows. Let $G_1 :=(E_1,V_1)=(E,V)$. Then, given $G_l$ for $1\leq l\leq L-1$, we define  a $(l+1)^{\text{th}}$ aggregated graph as $G_{l+1}=(E_{l+1},V_{l+1})$ where $V_{l+1}:=\{v^{(l+1)}_1,\ldots,v^{(l+1)}_{N_{l+1}}\}$ is a partition of $V_l$ and $E_{l+1}$ is computed thanks to $G_l$ as the following aggregation step:
\begin{enumerate}
\item if $i\not=j$, the edge between $v^{(l+1)}_i$ and $v^{(l+1)}_j$ in $G_{l+1}$ is equal to the sum of all edges between nodes of $G_l$ contained in $v^{(l+1)}_i$ and nodes of $G_l$ contained in $v^{(l+1)}_j$,
\item if $i=j$, loop $i$ in $E_{l+1}$ is equal to the sum of all edges between nodes of $G_l$ contained in $v^{(l+1)}_i$.
\end{enumerate}
Moreover, We denote by $A^{(l)}\in\mathcal{M}_{|V_{l}|}(\R)$ the corresponding symmetric adjacency matrix where entry $A^{(l)}_{ij}$ denotes the weight assigned between vertices $v_{i}^{(l)}$ and $v_{j}^{(l)}$ in $G_{l}$. The degree of a node $i$ is denoted $k^{(l)}_i$ and $m_l:=|E_l|=\frac{1}{2}\sum_i k^{(l)}_i$. We call $\bc_{l}\in \C_{l}$ a \textit{coloration} of level $l$ of graph $G_l$ any partition $\bc_{l}=\{c_{1,l},\ldots,c_{k,l}\}$ of $V_l$ where for any $i=1,\ldots, k$, $c_{i,l}\subseteq V_l$ is a set of nodes of $G_l$. Moreover, $\bc_{l}(i)$ denotes the community of vertex $i$ based on $\bc_{l}$. Moreover, we denote by $\map^{\bc_l}:V^{(l)}\to V^{(l+1)}$ the mapping of all nodes of $G_l$ in $G_{l+1}$ that groups all nodes of a same community according to $\bc_l$ in a single node of $G_{l+1}$. For instance, if for some $i$ we have $c_{i,l}=\{v_{1}^{(l)},\ldots, v_{r}^{(l)}\}$, then $\map^{\bc_l}(v)=\map^{\bc_l}(v')$ for any $v,v'\in c_{i,l}$.

 Finally, the decision to find the community of $i\in V$ thanks to $(C_l,G_l)_{l=1}^L$ is made of the following computation:
$$
\bc(i):=\bc_L\left(\map^{\bc_{L-1}}\circ\cdots\circ \map^{\bc_1}(i)\right),
$$ 
where $C_L(v)$ stands for the community of $v\in V^{(L)}$. 

Under these notations, we can define the family of priors $(p^{(l)}(\cdot|\bc_{l},G_l))_{l=1}^L$ where each $p^{(l)}(\cdot|\bc_{l},G_l)$ is defined in \eqref{improvedproposal} and acts on the graph of level $l$ from $l=1$ (the original graph) to $l=L$ (the highest level of aggregation). Endowed with this family of hierarchical priors, the proposal distribution is defined on $\mathcal{P}(\otimes_{l=1}^L \C_{l})$ as follows:
\begin{eqnarray}
\label{hierarchicalprior}
p(\bc'|\bc)=\sum_{l=1}^L\alpha_l p^{(l)}(\bc'_l|\bc_{l},G_l),\text{for any }\bc'=(\bc_1,\ldots, \bc_l)\in\otimes_{l=1}^L \C_{l},
\end{eqnarray}
where $\sum_{l}\alpha_l=1$ whereas $\bc=(\bc_1,\ldots, \bc_L)$ and $\bc'=(\bc'_1, \ldots, \bc'_{L})$ contain colorations of graphs at different levels.
\medskip

The principle of this new MH algorithm is illustrated in \autoref{algohmh} and maintain the family $(C_l,G_l)_{l=1}^{L}$ thanks to \eqref{hierarchicalprior}.

\begin{algorithm}
\caption{Hierarchical MH Community Detection}
\label{algohmh}
\begin{algorithmic}[1]
  \scriptsize
  \STATE Initialization $\lambda>0$, $L\geq 1$, $(G_l^{(0)},\bc_l^{(0)})_{l=1}^L$.
  \STATE For $k=1, \ldots, N$:
  \STATE Draw $\bc'\sim p$ where $p(\cdot|\bc_l^{(k-1)},G_l^{(k-1)})\in\PP(\otimes_{l=1}^L \C_{l})$ is defined in \eqref{hierarchicalprior}.
  \STATE If $\bc'$ has been proposed by the $l^{\text{th}}$ prior $p^{(l)}$ for some $l=1, \ldots,L$ update $\bc^{(k)}=\bc'$ with Metropolis ratio :
  \begin{align}
\label{ratio2}
\rho = 1\wedge \left(r_{\bc^{(k-1)} \rightarrow \bc'}\frac{\exp\left(\lambda Q^{\bc'_l}\right)}{\exp\left(\lambda Q^{\bc_l^{(k-1)}}\right)}\right),\mbox{ where }r_{\bc^{(k-1)} \rightarrow \bc'}:=p(\bc_l^{(k-1)}|\bc'_l)/p(\bc'_l|\bc_l^{(k-1)}).
	\end{align}
	\STATE If $\bc'$ has been accepted and has used the $l^{\text{th}}$ prior $p^{(l)}$ for some $l=1,\ldots,L$, maintain $(G_{k'})_{k'=l+1}^L$ as follows: 
	\STATE For $k'=l,\dots, L-1$ 
	\STATE Update $V_{k'+1}$ thanks to $\map^{\bc_{k'}^{(k)}}$,
	\STATE Update $E_{k'+1}$ thanks to the aggregation step define above.
\end{algorithmic}
\end{algorithm}

In 5: above, if $\bc'$ has used the $l^{\text{th}}$ prior $p^{(l)}$ for some $l=1,\ldots,L$, we update graphs $G^{(k)}_{k'}$, $k'=l+1,\ldots, L$ as follows:
\begin{enumerate}
\item if $i$ and $j$ belong to different communities in $C^{(k)}_{l}$, $i$ is re-mapped to the same node than $j$ in $G^{(k)}_{k'}$, $k'=l+1,\ldots, L$.
\item if $i$ and $j$ belong to the same community in $C^{(k)}_{l}$, $i$ is re-mapped to a new single node community in $G^{(k)}_{k'}$, $k'=l+1,\ldots, L$.
\end{enumerate}


\subsection{Modularity gain}
In \autoref{algohmh}, several states of proposals $\bc'$ lead to the same coloration $\{\bc'(i),i\in V_1\}$. The choice of $\bc'$ based on  $p^{(l)}$ may lead to the situation where $\bc'_k = \bc_k$ for $k=l^*,\ldots,L$ for some $l<l^*\leq L$ and this results in $\bc(i)=\bc'(i)$ for any $i\in V_1$. For instance this happens if $\bc_k$ is a single community that contains all nodes for $k=l^*, \ldots,L$ and if the proposal at level $l$ doesn't create a new community.

As a consequence, if for $\bc=(\bc_1,\ldots, \bc_L)$ and $\bc'=(\bc'_1, \ldots, \bc'_{L})$ we have $\bc_L=\bc'_L$ , the real modularity gain for this move is actually 0.

If $l = L$ then \ref{eq:delta_plus}, \ref{eq:delta_minus} and \ref{deltamodularity} apply.

If $l < L$ and $\bc_L \neq \bc'_L$ the modularity change by moving node $i$ is given by the following formulas.

First, let us introduce the following notation :
\begin{align*}
&\bc_{L,l}(i):=\bc_L\left(\map^{\bc_{L-1}}\circ\cdots\circ \map^{\bc_l}(i)\right)
\end{align*}

If $l < L$ and $\bc_L \neq \bc'_L$ the modularity gain by moving node $i$ is given by :
\begin{itemize}
\item If node $i$ joins an existing community at level $l$ then at level $L$ the modularity held by the community $c=C'_{L,l}(i)$ will be modified by the following quantity : 
\begin{align*}
\Delta Q_+^{C\rightarrow C'} = &\frac{1}{m}\left(\left(\sum_{j \in V}  A_{ij}^{(l)} \delta\left(C_{L,l}(j),c\right)\right)-\frac{k_i^{(l)}\left(k_{c}^{C_L}+\frac{1}{2}k_i^{(l)}\right)}{2m}\right).
\end{align*}
The same formula applies if $i$ joins a new single node community at level $l$ which implies the creation of a new node in its own community $c$ at level $L$.
\item When $i$ leaves its community at level $l$, the modularity held by the community $C_{L,l}(i)$ will be modified by the following quantity :
\begin{align*}
\Delta Q_-^{C\rightarrow C'} = &-\frac{1}{m}\left(\left(\sum_{j \in V}  A_{ij}^{(l)} \delta\left(C_{L,l}(j),C_{L,l}(i)\right)\right)-\frac{k_i^{(l)}\left(k_{C_{L,l}(i)}^{C_L}-\frac{1}{2}k_i^{(l)}\right)}{2m}\right)
\end{align*}
\end{itemize}
Finally we use the formula $\Delta Q^{C\rightarrow C'} = \Delta Q_+^{C\rightarrow C'} + \Delta Q_-^{C\rightarrow C'}$.

\subsection{Conditional probabilities}
The calculus of $r_{\bc\to\bc'}$ in \eqref{ratio2} follows Section \ref{sec:3}. The only difference is the hierarchical prior defined in \eqref{hierarchicalprior}. Then, given a state $\bc=(\bc_1, \ldots, \bc_L)$ and a proposal $\bc'=(\bc'_1,\ldots, \bc'_L)$, the probability to come back is given by:
$$
p(\bc|\bc')=\alpha_{l'}p^{(l')}(\bc_{l'}|\bc'_{l'},G_{l'}),
$$
where $l'$ is the level chosen by $\bc'$ and $p^{(l')}(\bc_{l'}|\bc'_{l'},G_{l'})$ is defined in \autoref{condproba} applied to the aggregated graph $G_{l'}$.

\section{Dynamic Metropolis Hasting graph clustering}
\label{sec:5}
The purpose of this section is to adapt the previous algorithm to the dynamic graph clustering problem. The challenge is to maintain a clustering for a sequence of graphs $(G_t)_{t\geq 1}$, where $G_t$ is derived from $G_{t-1}$ by applying a small number of local changes. This problem has been proven to be NP-hard (see\cite{maillard}) and several authors has tried to propose dynamic clustering algorithms. \cite{Gorke2009} proposes a dynamization of the minimum-cut trees algorithm and allows to keep consecutive clustering similar. More recently,  \cite{maillard} proposes heuristics for dynamization of greedy search algorithms of \cite{Brandes} and \cite{blondel08}, where at each time $t$, a so-called preclustering decision is passed to the static algorithm.
\medskip

In a prediction framework, \cite{rakhlintuto} proposes to derive online nodes classification algorithms based on a general minimax analysis of the so-called regret (see also \cite{rakhlin2012relax}). In this problem, nodes have associated labels that could be correlated with the topology of the graph edges. Several techniques are proposed, based on a convex relaxation of the problem and also the use of surrogate losses (see \cite{rakhlinsurrogate}). 
\medskip

\autoref{onlinehmh} describes the online procedure. Coarselly speaking, the principle of the algorithm is to run at each new observation $t$ \autoref{algohmh} from the endpoint of step $t-1$. The choice of $N(t)$ depends on the frequency of the sequence $(G_t)_{t\geq 1}$ and the execution time of \autoref{algohmh}. We recommend to run \autoref{algohmh} until a new incoming observation arrives. 

\begin{algorithm}[H]
\caption{Online MH Community Detection}
\label{onlinehmh}
\begin{algorithmic}[1]
  \scriptsize
  \STATE Initialization $\lambda>0$, $L\geq 1$, $(G_l^{(0,0)},C_l^{(0,0)})_{l=1}^L$,$N(0)=0$.
  \STATE For $t=1,\ldots, T$:
  \STATE $(\bc^{(t,0)},G^{(t,0)}):=(\bc^{(t-1,N(t-1))},G^{(t-1,N(t-1))})$
  \STATE For $k=1, \ldots, N(t)$:
  \STATE Draw $\bc'\sim p$ where $p(\cdot|C_l^{(t,k-1)},G_l^{(t)})\in\PP(\otimes_{l=1}^L \C_{l})$ is defined in \eqref{hierarchicalprior}.
  \STATE If $\bc'$ has been proposed by the $l^{\text{th}}$ prior $p^{(l)}$ for some $l=1, \ldots,L$ update $\bc^{(t,k)}=\bc'$ with Metropolis ratio :
  \begin{align}
\label{ratio2}
\rho = 1\wedge \left(r_{\bc^{(t,k-1)} \rightarrow \bc'}\frac{\exp\left(\lambda Q^{\bc'_l}\right)}{\exp\left(\lambda Q^{\bc_l^{(t,k-1)}}\right)}\right),\mbox{ where }r_{\bc^{(t,k-1)} \rightarrow \bc'}:=p(\bc_l^{(t,k-1)}|\bc'_l)/p(\bc'_l|\bc_l^{(t,k-1)}).
	\end{align}
	\STATE If $\bc'$ has been accepted and has used the $l^{\text{th}}$ prior $p^{(l)}$ for some $l=1, \to,L$, maintain $(G_{k'})_{k'=l+1}^L$ as follows: 
	\STATE For $k'=l,\dots, L-1$  
	\STATE Update $V_{k'+1}$ thanks to $\map^{\bc_{k'}^{(t,k)}}$,
	\STATE Update $E_{k'+1}$ thanks to the aggregation step define above.
\end{algorithmic}
\end{algorithm}
In 7: above, if $\bc'$ has used the $l^{\text{th}}$ prior $p^{(l)}$ for some $l=1, \to,L$, we update graphs $G^{(t,k)}_{k'}$, $k'=l+1,\ldots, L$ as follows:
\begin{enumerate}
\item if $i$ and $j$ belong to different communities in $\bc^{(t,k)}_{l}$, $i$ is re-mapped to the same node than $j$ in $G^{(t,k)}_{k'}$, $k'=l+1,\ldots, L$.
\item if $i$ and $j$ belong to the same community in $\bc^{(t,k)}_{l}$, $i$ is re-mapped to a new single node community in $G^{(t,k)}_{k'}$, $k'=l+1,\ldots, L$.
\end{enumerate}
\section{Conclusion and Acknowledge}
\label{sec:6}
In this paper, we propose an online community detection algorithm based on a dynamic optimization of the modularity. Our method appears to be a dynamization of \cite{blondel08} and uses a Metropolis Hasting formulation. 
\medskip

In a batch setting, the given algorithm shows practical results comparable to the so-called Louvain algorithm introduced in \cite{blondel08}. However, we omit arbitrary decisions such as the time to aggregate, the number of aggregations or the order of observations of the nodes in the first pass of Louvain. More precisely, we introduce a well-chosen instrumental measure in the MH paradigm in order to include aggregation in the proposal mooves. A precise calculation of the conditional probabilities, as well as modularity deviations, allows to construct a suitable Markov chain with ergodic properties.
\medskip

Finally, this MCMC version of Louvain allows to build coarselly an online version where communities are constructed in an online fashion. Experiments over simulated graphs (such as preferential attachment models PA($n$), or preferential attachment model with seeds, see \cite{bubeck2015influence}) as well as real-world graph databases ( see \href{http://konect.uni-koblenz.de/}{The Koblenz Network Collection
} for instance) are in progress. A dynamic and interactive visualization using a recent javascript framework D3.js is also coming up to illustrate the output of the algorithm in real time. 
\medskip

This work falls into a project of open innovation between two french startups Fluent Data and Artfact, and is supported by the French government under the label Jeune Entreprise Innovante (J.E.I.).

\bibliographystyle{plain}
\bibliography{reference}
\end{document}